\documentclass[twocolumn, 10pt]{article}
\usepackage{amsmath,amssymb}
\usepackage{bm}
\usepackage[dvipdfmx]{graphicx,hyperref}
\usepackage[format=hang,labelfont=bf,textfont=small,singlelinecheck=false,justification=raggedright,margin={12pt,12pt},figurename=Fig.]{caption}
\usepackage{titlesec}
\usepackage{textcomp}
\usepackage[dvipsnames]{xcolor}
\usepackage{cite}

\makeatletter
\def\affiliation#1{\gdef\@affiliation{#1}}
\def\abstract#1{\gdef\@abstract{#1}}
\def\graphabst#1{\gdef\@graphabst{#1}}
\def\keywords#1{\gdef\@keywords{#1}}
\def\corresp#1{\gdef\@corresp{#1}}

\newcommand{\MakeTitle}{
  \newpage
  \null
  \vskip 2em%
  \begin{center}%
  \Large \@title\par
  \vskip 1em%
  \large \@author
  \end{center}
  \noindent\@affiliation\par
  \vskip 1em%
  \noindent\@corresp\par
  \vskip 1em%
  \noindent\@abstract\par
  \noindent\@graphabst\par
  \vskip 1em%
  \noindent\@keywords\par
}
\makeatother

\newcommand*{\TitleFont}{%
      \usefont{\encodingdefault}{\rmdefault}{}{n}%
      \fontsize{18}{12}%
      \selectfont}

\titleformat{\section}
  {\normalfont\fontsize{10}{11}\bfseries}{\thesection.}{2pt}{}
  \titlespacing*{\section}{0pt}{12pt}{6pt}

\titleformat{\subsection}
  {\normalfont\fontsize{10}{10}\bfseries}{\thesubsection.}{2pt}{}
  \titlespacing*{\subsection}{0pt}{6pt}{0pt}
  
\titleformat{\subsubsection}
  {\normalfont\fontsize{10}{10}\bfseries}{\thesubsubsection.}{2pt}{}
  \titlespacing*{\subsubsection}{0pt}{6pt}{0pt}
  
\normalsize

\title{\TitleFont Drag of an elliptic intruder\\
in a two-dimensional granular environment}
\author{Takumi KUBOTA$^{1 }$, Haruto ISHIKAWA$^{1 }$, Satoshi TAKADA$^{1,2\dagger}$}
\affiliation{$^{1}$Department of Mechanical Systems Engineering, 
    Tokyo University of Agriculture and Technology,
    2--24--16 Nakacho, Koganei, Tokyo
    184--8588 Japan, \\
    $^{2}$Institute of Engineering, 
    Tokyo University of Agriculture and Technology,
    2--24--16 Nakacho, Koganei, Tokyo 
    184--8588 Japan}
\corresp{$^{\dagger}$Corresponding author, 
    takada@go.tuat.ac.jp, 
    Tel.:$+$81-42-388-7702}

\abstract{\textbf{Abstract}: The drag of an elliptic intruder in a two-dimensional granular environment is numerically studied.
The movement parallel to the major axis of the intruder is found to be unstable.
The drag law is given by the sum of the yield force and the dynamic term, the latter of which is approximately reproduced by a simple collision model.
The flow field around the intruder for sufficiently larger drag force is well fitted by the streamlines obtained from the perfect fluid.
The stress fields around the intruder are also investigated when the movement of the intruder is balanced with interactions with the surrounding particles.
The Airy stress function is found to well reproduce the stress fields once the stress on the surface of the intruder is given.}

\graphabst{
\begin{center}
\end{center}
}

\begin{document}

\onecolumn
\MakeTitle

\twocolumn
%%%%%%%%%%%%%%%%%%%%%%%%%%%%%%%%%%%
\section{Introduction}
It is important to know the resistance of granular materials.
We know that if the drag force acting on an object is less than a certain threshold, the object can not move due to the resistance from the surrounding particles and remains stationary.
On the other hand, the object can move if the force is large.
By knowing the resistance from the surrounding particles, we can estimate the minimum force to keep the object moving and expect to reduce the energy loss.

The drag in granular environments is known to be different from that in fluids.
The relationship between the resistance and the velocity varies with the density of granular materials.
Takehara and Okmura\cite{Takehara14} reported that the drag law is expressed as the sum of yield force and resistance proportional to the square of the velocity from experiments.
In this way, the drag in granular materials is studied in many setups and situations.
However, most of the previous papers focus on the drag of an idealistic intruder, that is, circular \cite{Cheng07,Cheng14,Sano12,Sano13,Takada17,Kubota21a,Bharadwaj06,Hilton13,Dhiman20,Takehara10,Takehara14,Wassgren03,Cruz16,Espinosa21,Kubota21b}, spherical \cite{Takada20,Takada20jet,Kumar17,Jewel18}, or cylinder \cite{Reddy11,Chang22,Guillard14,Pal21,Hossain20Mobility,Hossain20Rate-dependent,Hossain20Drag} in two- and three-dimensional systems, respectively \cite{Tripura21,Escalante17,Dziugys01, Carvalho22}.
Little is known about more general shapes such that the drag law changes when the shape is a triangle.
Tripura {\it et al.}\ reported that the drag force is independent of the intruder shape if the cross-sectional area of an object in the drag direction is the same\cite{Tripura21}.
However, they focus only on the case where the object is in steady motion, and there have been few studies in the region where the object can not move by the surrounding particles.

In previous studies\cite{Kubota21a,Kubota21b}, we have analyzed drag simulations in the case of one or two circular intruders as simple systems.
For the case of one intruder above the threshold, it was found that the drag force can be reproduced by a collision model.
Below the threshold, the stress fields in front of the intruder are well reproduced by applying the theory of elasticity \cite{Kubota21a}.

However, the circular shape is an ideal shape, and more complex shapes need to be considered for applications.
Therefore, as a first step, we consider an elliptic shape.
As mentioned above, some facts have already been reported \cite{Tripura21} in the case above the threshold, that is, the regime where intruders move but are not clear below the threshold of the drag force by discrete element method (DEM) \cite{Cundall}.
To this end, we investigate the drag of an elliptic intruder in a wide range of the drag force.
We report the existence of a threshold, and above and below which, we investigate the drag law, the streamlines, and the applicability of the theory of elasticity.

The organization of this paper is as follows:
In the next section, the model and setup of our study are briefly explained.
The section \ref{sec:results} is the main part of this paper, where we present the drag law and stress fields around the intruder.
In Sect.\ \ref{sec:conclusion}, we discuss and conclude our results.
In Appendix \ref{sec:coll_detection}, we explain how to derive the contact forces between the elliptic intruder and the surrounding particles.
In Appendix \ref{sec:viscous_elliptic}, we explain the representation of the stream function in a two-dimensional viscous fluid.
In Appendix \ref{sec:elliptic_stressfield}, the derivation of the stress fields around the elliptic intruder is presented.

%%%%%%%%%%%%%%%%%%%%%%%%%%%%%%%%%%%
\section{Model and setup}\label{sec:model}
\begin{figure}[htbp]
    \centering
    \includegraphics[width=0.8\linewidth]{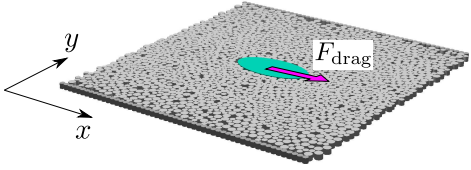}
    \caption{Setup of our simulation.
    An elliptic intruder is set at the origin at initial.
    The drag force acts on the intruder in the $x$ direction.
    The angle $\theta$ is introduced as that of the major axis of the intruder from the $x$ axis.
    Here, we use $\varphi=0.80$, $2R_1=15d$, and $2R_2=7.5d$.}
    \label{fig:setup}
\end{figure}

In this section, let us explain our model and setup of simulations.
As shown in Fig.\ \ref{fig:setup}, we first put an elliptic intruder whose lengths of the major and the minor axes are $2R_1$ and $2R_2$, respectively and the mass is $M$.
We introduce the angle of the major axis from the $x$ direction as $\theta$.
Next, bidisperse particles are distributed in the system.
We choose the diameter and mass of smaller particles as $d$ and $m$, respectively.
We assume that the larger ones consist of the same material as the smaller ones, and the diameter and mass are given by $1.4d$ and $1.4^2m$, respectively.
We note that the mass of intruder $M$ is calculate by $M/m = 4R_1R_2/d^2$.
We adopt the periodic boundary condition in the $x$ direction and put smaller particles at $y=\pm L_y/2$ as bumpy boundaries \cite{Takada17, Kubota21a, Kubota21b}.

%%%%%%%%%%%%%%%%%%%%%%%%%%%%%%%%%%%
%%%%%%%%%%%%%%%%%%%%%%%%%%%%%%%%%%%

Here, the procedure to detect collisions between the intruder and the surrounding particles is explained in Appendix \ref{sec:coll_detection}.

We perform the DEM to observe time evolution of the system.
The equation of translational motion of the particles is described by
\begin{equation}
    m_i\frac{\mathrm{d}^2\bm{r}_i}{\mathrm{d}t^2} = \sum_j \bm{F}_{ij} + F_{\rm drag}\hat{\bm{e}}_x\delta_{0,i} -\mu_{\rm b}g\hat{{\bm v}}_i\left(1-\delta_{0,i}\right),
\end{equation}
where $i=0$ and $i \ge 1$ indicate an intruder and surrounding particles, respectively.
Here, $\bm{r}_i$ is the position vectors of $i$-th particle, $\bm{F}_{ij}$ is the contact force between $i$-th and $j$-th particles, $F_{\rm drag}$ is the drag force acting on the intruder, $\hat{\bm{e}}_x$ is the unit vector parallel to $x$ direction, $\delta_{i,j}$ is the Kronecker delta, $\mu_{\rm b}$ is the coefficient of the basal friction, $g$ is the acceleration of gravity, and $\hat{\bm{v}}_i$ is the unit vector parallel to the velocity of $i$-th particle \cite{Luding08}.

The contact force $\bm{F}_{ij}$ is described by $\bm{F}_{ij} = F_{ij}^n\bm{n}+ F_{ij}^t\bm{t}$ with the magnitude of the normal and tangential force $F_{ij}^n$ and $F_{ij}^t$ , respectively and the unit vectors for the normal and tangential direction $\bm{n}$ and $\bm{t}$, respectively.
The normal force $F_{ij}^n$ is given by
\begin{equation}
    F_{ij}^n = k_n\delta - \eta_n\left(\bm{v}_{ij}\cdot \hat{\bm{r}}_{ij}\right) \Theta\left(d_{ij}-r_{ij}\right),
\end{equation}
where $k_n$ and $\eta_n$ are the spring constant and the damping coefficient in the normal direction, respectively, $\delta$ is the overlap between contacting two particles, $\bm{v}_{ij} = \bm{v}_{i} - \bm{v}_{j}$ is the relative velocity, $r_{ij}=|\bm{r}_i-\bm{r}_j|$, $\hat{\bm{r}}_{ij}=\bm{r}_{ij}/r_{ij}$, $\Theta(x)$ is the step function, and $d_{ij}=(d_i+d_j)/2$.
The tangential force $F_{ij}^t$ is given by
\begin{equation}
    F_{ij}^t = \min (\mu F_{ij}^n, \left| -k_t\xi - \eta_t v_{ij}^t\right|) \Theta\left(d_{ij}-r_{ij}\right),
\end{equation}
where $\mu$ is the coefficient of the Coulombmic friction between contacting particles, $k_t$ and $\eta_t$ are the spring constant and the damping coefficient in the tangential direction, respectively, $\xi$ is the displacement in the tangential direction, and $v_{ij}^t = |\bm{v}_{ij}-(\bm{v}_{ij}\cdot \hat{\bm{r}}_{ij})\hat{\bm{r}}_{ij}|$ \cite{Luding08}.

The equation of motion for rotational motion of the particles is described by
\begin{equation}
    I_i\frac{\mathrm{d}^2\theta_i}{\mathrm{d}t^2} = T_i,
\end{equation}
where $I_0 = M(R_1^2+R_2^2)/4$, $I_i=m_id_i^2/8$ ($i \ge 1$) is the moment of inertia of $i$-th particle, $\theta_i$ is the angle of $i$-th particle, $T_i$ is the torque applied to $i$-th particle.

In the simulation, we choose $2R_1=15d$, $2R_2=7.5d$, the packing fraction $\varphi=0.80$, the number of the surrounding particles $N=10000$, $\mu=\mu_{\rm b}=0.20$, $k_n=k_t$, $\eta_n=\eta_t$, and $g=1.0\times 10^{-4}k_nd/m$ in most cases.

%%%%%%%%%%%%%%%%%%%%%%%%%%%%%%%%%%%
\section{Results}\label{sec:results}
In this section, we present our main results obtained from the simulations.
In the first subsection, we focus on the relationship between the drag force $F_{\rm drag}$ and the steady speed $V$ of the intruder.
In the next subsection, the stress fields around the intruder are studied from the information on the surface of the intruder.

%%%%%%%%%%%%%%%%%%%%%%%%%%%%%%%%%%%
\subsection{Drag law}

%%%%%%%%%%%%%%%%%%%%%%%%%%%%%%%%%%%
\begin{figure}[htbp]
    \centering
    \includegraphics[width=\linewidth]{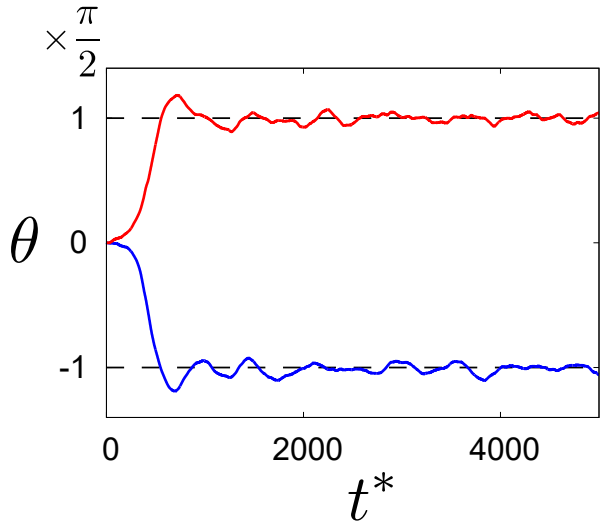}
    \caption{Typical time evolution of the angle of the intruder $\theta$ for $\varphi=0.80$, $2R_1=15d$, $2R_2=7.5d$, and $F_{\rm drag}=1.0\times10^{-1}k_nd$, where we have introduced $t^*\equiv t\sqrt{k_n/m}$.}
    \label{fig:theta}
\end{figure}
%%%%%%%%%%%%%%%%%%%%%%%%%%%%%%%%%%%
As shown in Fig.\ \ref{fig:theta}, the movement parallel to the major axis is unstable.
Even if the initial direction of the major axis of the intruder $\theta$ is parallel to the drag direction, this direction tends to be perpendicular to the drag direction ($\theta=\pi/2$ or $-\pi/2$ depending on the initial configuration).
This can be understood as following:
If there exists a small perturbation $\Delta \theta$, the torque works on the intruder with the same sign of $\Delta \theta$, which means that the small perturbation enhances the rotation of the intruder for $\theta\simeq 0$.
Therefore, from now on, we present results when we pull the intruder with the initial angle $\theta=\pi/2$.

\begin{figure}[htbp]
    \centering
    \includegraphics[width=\linewidth]{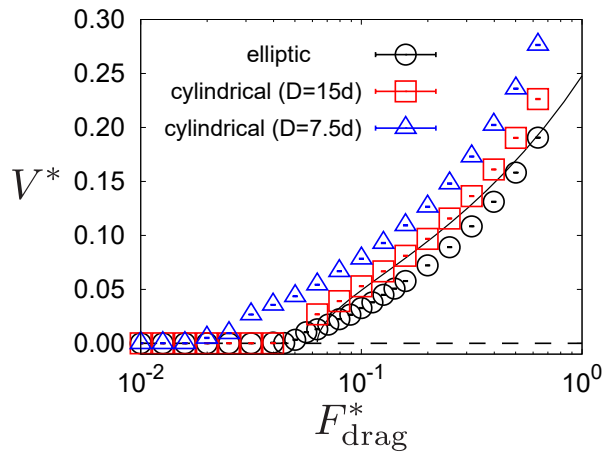}
    \caption{Relationship between the drag force $F_{\rm drag}$ and the steady speed $V$ for $\varphi=0.80$, $2R_1=15d$, and $2R_2=7.5d$.}
    \label{fig:FV}
\end{figure}

Figure \ref{fig:FV} shows the relationship between the drag force $F_{\rm drag}$ and the steady speed $V$
When we compare the results of the elliptic case and the cylindrical case with $D=15d$, the resistance becomes larger for the elliptic case.
This can be understood as follows:
If the surrounding particle collides with the intruder at a certain $y_0$, the corresponding $x$ coordinates become $R_2\sqrt{1-y_0^2/R_1^2}$ and $\sqrt{R^2-y_0^2}$ for the elliptic and cylindrical cases, respectively, where $R=D/2$ is the radius of the cylinder.
If we put the angle between the $x$ direction and the normal vector at the colliding point as $\beta$ the detailed derivation is given in the next subsection, and the work from the particle to the intruder is proportional to $\cos^2\beta$ \cite{Kubota21b, Garzo_book}.
Then, this angle becomes smaller for the elliptic case, which also means that $\cos\beta$ becomes larger.
Summing up all the contributions around the surface of the intruder, the drag resistance becomes larger for the elliptic case.

%%%%%%%%%%%%%%%%%%%%%%%%%%%%%%%%%%%
%%%%%%%%%%%%%%%%%%%%%%%%%%%%%%%%%%%

%%%%%%%%%%%%%%%%%%%%%%%%%%%%%%%%%%%
\subsection{Collision model for $F_{\rm drag}\gtrsim F_{\rm Y}$}
Let us derive the relationship between the dynamic part of the drag force $F_{\rm dyn}$ and the steady speed $V$.
We consider the coordinate that the intruder is set at the origin with the angle $\theta=\pi/2$.
Moreover, we consider a situation where the surrounding particles move towards the negative $x$ direction with the steady speed $V$ and collide with the intruder.
First, we consider a point $(R_2^\prime\cos\vartheta, R_1^\prime\sin\vartheta)$ ($-\pi/2\le \vartheta\le \pi/2$), where we have introduced $R_1^\prime \equiv R_1+d/2$ and $R_2^\prime \equiv R_2+d/2$.
This point indicates the center of the colliding particle whose diameter is $d$.
Here, this system can be regarded as the collision between the mass point and the elliptic intruder with $R_1^\prime$ and $R_2^\prime$.
Then, the unit normal vector at this point becomes
\begin{equation}
    \bm{n}
    =(\cos\beta, \sin\beta)^T,
\end{equation}
with
\begin{subequations}
\begin{align}
    \cos\beta &= \frac{R_1^\prime\cos\vartheta}{\sqrt{R_1^{\prime2}\cos^2\vartheta + R_2^{\prime2}\sin^2\vartheta}},\\
    \sin\beta &= \frac{R_2^\prime\sin\vartheta}{\sqrt{R_1^{\prime2}\cos^2\vartheta + R_2^{\prime2}\sin^2\vartheta}}.
\end{align}
\end{subequations}
It should be noted that the angle of the vector $\bm{n}$ from the $x$ axis is not $\vartheta$ but $\beta$.
The momentum change in the $x$ direction due to this collision is given by $\Delta p_x \simeq \{[(2/3)+e_n-e_t/3]\cos^2\beta + (1+e_t)/3\}m_{\rm r}V$ with the reduced mass $m_{\rm r}\equiv Mm/(M+m)$ \cite{Garzo_book}.
In addition, the volume of the collision cylinder corresponding to the parameter $\vartheta$ becomes $\mathcal{V}_{\rm coll}= V(R_1^{\prime2}\cos^2\vartheta + R_2^{\prime2}\sin^2\vartheta)^{1/2}\cos\beta$.
Integrating the momentum change in the $x$ direction over the surface of the (modified) intruder, we get
\begin{equation}
    F_{\rm dyn}
    =\int_{-\pi/2}^{\pi/2} n \Delta p_x \mathcal{V}_{\rm coll}d\vartheta
    \simeq \alpha V^2,
    \label{eq:F_V}
\end{equation}
with
\begin{align}
    &\alpha
    \equiv \frac{4}{\pi}\varphi m_{\rm r}\frac{2R_1+d}{d^2}\nonumber\\
    &\times\left[\frac{2+3e_n-e_t}{3}
    \left(\frac{1}{\epsilon^2}-\frac{1+\epsilon^2}{2\epsilon^3}\log \frac{1+\epsilon}{1-\epsilon} \right)+\frac{1+e_t}{3}\right],
\end{align}
where we have introduced the eccentricity $\epsilon\equiv[1-(2R_2+d)^2/(2R_1+d)^2]^{1/2}$.
In the above derivation, we regard all particles around the intruder as monodisperse with the diameter $d$ and the mass $m$.
This simplification does not affect the results.
We note that we assume that the momentum arm is fixed as a constant independent of the collision point.
Our collision model \eqref{eq:F_V} recovers that for a cylindrical intruder \cite{Kubota21a} if we take the limit $R_2\to R_1$.
This estimation works well as shown in Fig.\ \ref{fig:FV} at least when the dimensionless speed $V^*$ is sufficiently smaller than unity.

%%%%%%%%%%%%%%%%%%%%%%%%%%%%%%%%%%%
\subsection{Streamlines around the intruder for $F_{\rm drag}\gtrsim F_{\rm Y}$}
\begin{figure}[htbp]
    \centering
    \includegraphics[width=\linewidth]{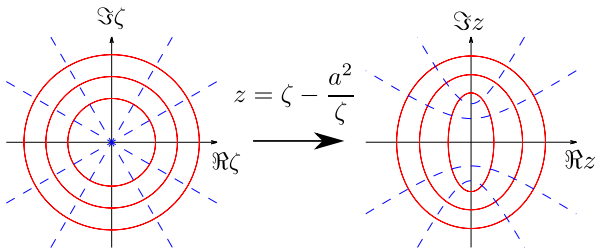}
    \caption{Conformal map \eqref{eq:mapping} from the polar coordinates ($\Re \zeta$, $\Im \zeta$) to the elliptic coordinates ($\Re z$, $\Im z$).}
    \label{fig:mapping}
\end{figure}

In this subsection, we compare streamlines obtained from the simulation with those of perfect fluid around an elliptic intruder.
Now, let us derive the latter from Ref.\ \cite{Turnbull10}.
In this case, it is well known that a complex velocity function $W(z)=\phi(z)+i \psi(z)$ is a powerful tool to describe a two-dimensional flow \cite{Lamb, Batchelor}, where $\phi(z)$ and $\psi(z)$ are the velocity potential and the stream function, respectively, with $z=x+iy$, and $i$ is an imaginary number.
First, we consider a flow around a cylinder.
Assuming a uniform flow with the speed $-V$ at infinity in the $x$-direction, the stream function can be described by \cite{Lamb}
\begin{equation}
    W(z) = -V\left(\zeta+\frac{c^2}{\zeta}\right).
    \label{eq:w_cylinder}
\end{equation}
When the $\zeta$-plane is isometrically mapped to the $z$-plate by
\begin{equation}
    z=\zeta-\frac{a^2}{\zeta},
    \label{eq:mapping}
\end{equation}
the flow around a cylinder \eqref{eq:w_cylinder} is transformed into the flow around an elliptical cylinder whose minor axis coincides with the flow ($x$-)direction.
Actually, in the $\zeta$-plane, the circle $\zeta = c e^{i\theta}$ of radius $c$ located at the origin is projected to
\begin{equation}
    z =\left(c-\frac{a^2}{c}\right)\cos\theta 
    + i\left(c+\frac{a^2}{c}\right)\sin\theta,
\end{equation}
in the $z$-plane (see Fig.\ \ref{fig:mapping}).
We note that $c \ge a$ should be satisfied.
Now, both $c$ and $a$ should be 
\begin{equation}
    c = \frac{R_1+R_2}{2},\quad
    a = \frac{\sqrt{R_1^2-R_2^2}}{2},
\end{equation}
respectively.
From the complex velocity function \eqref{eq:w_cylinder} with the mapping \eqref{eq:mapping}, we can obtain the complex velocity function which describes the flow around an elliptic intruder as
\begin{equation}
    W(z) = \frac{Vz}{R_1-R_2}\left(R_2- R_1\sqrt{1+\frac{R_1^2-R_2^2}{z^2}}\right).
    \label{eq:w_elliptic}
\end{equation}
Then, the streamlines are characterized by $\psi(z)= \Im W(z) = {\rm const}$ \cite{Lamb}.

%%%%%%%%%%%%%%%%%%%%%%%%%%%%%%%%%%%
%%%%%%%%%%%%%%%%%%%%%%%%%%%%%%%%%%%

\begin{figure}[htbp]
    \centering
    \includegraphics[width=0.75\linewidth]{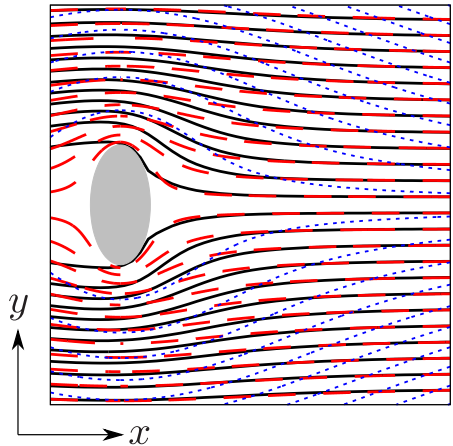}
    \caption{Streamlines obtained from the simulation (solid lines) and those of the perfect fluid (dashed lines) for $2R_1=15d$ and $2R_2=7.5d$.
    The streamlines of the viscous fluid (dotted lines) are also plotted.
    The parameters used in the simulation are $\varphi=0.80$ and $F_{\rm drag}=1.0\times10^{-1} k_n d$.}
    \label{fig:streamlines}
\end{figure}

Figure \ref{fig:streamlines} exhibits the comparison of the streamlines from the simulation with those of the perfect fluid given by Eq.\ \eqref{eq:w_elliptic}, where we also plot those from the viscous fluid which is explained in the next paragraph.
The theoretical streamlines well reproduce the simulation results in front of the intruder.
Now, we note that the streamlines from the simulations are inconsistent with the assumption which is used when we derive the collision model, where the trajectories of a particle colliding with the intruder are straightforward.
Actually, this assumption is not right because particles collide with each other before they collide with the intruder.
However, this assumption works well as a zeroth-order approximation to derive the drag law because the collision model works well as compared to the simulation results.
%%%%%%%%%%%%%%%%%%%%%%%%%%%%%%%%%%%
%%%%%%%%%%%%%%%%%%%%%%%%%%%%%%%%%%%

We also compare the results with streamlines of the viscous fluid.
It is well known that the stream function $\Psi$ of the viscous fluid satisfies the biharmonic equation $\Delta \Delta \Psi = 0$.
Following the procedure reported in Ref.\ \cite{Raynor02}, we can write the solution of the biharmonic equation as
\begin{align}
    \Psi
    &= \left[\mathcal{A}\cosh \varrho + \mathcal{B}\sin\varrho 
    + \mathcal{C}\varrho\cosh\varrho\right.\nonumber\\
    &\hspace{1em}\left.+ \mathcal{D}\left(\cosh^3\varrho + 6\varrho^2 \cosh\varrho 
    - 15\varrho \sinh\varrho\right)\right]\cos\vartheta,
    \label{eq:Psi_viscous}
\end{align}
where the expressions of the coefficients $\mathcal{A}$, $\mathcal{B}$, $\mathcal{C}$, and $\mathcal{D}$ are listed in Table \ref{fig:KABCD} (see Appendix \ref{sec:viscous_elliptic} for details).
As shown in Fig.\ \ref{fig:streamlines}, these streamlines cannot capture the behavior near the surface of the intruder, where the streamlines are bent further from those of the perfect fluid.

%%%%%%%%%%%%%%%%%%%%%%%%%%%%%%%%%%%
\begin{table*}[htbp]
    \caption{The expressions of $\mathcal{K}$, $\mathcal{A}$, $\mathcal{B}$, $\mathcal{C}$, and $\mathcal{D}$ \cite{Raynor02}.}
    \label{fig:KABCD}
    \centering
    \begin{tabular}{ll}\hline
    $\displaystyle \mathcal{K}=$
    & $\displaystyle \varrho_1-\varrho_0 
    -\frac{1}{4}\frac{(\cosh^2\varrho_1-\cosh^2\varrho_0) + 6(\varrho_1^2-\varrho_0^2)-15(\varrho_1\tanh\varrho_1-\varrho_0\tanh\varrho_0)}{\cosh\varrho_1 \sinh\varrho_1 - 2\coth\varrho_1+ 3\varrho_1}$\\
     & $\displaystyle -\cosh^2\varrho_0(\tanh\varrho_1-\tanh\varrho_0)\left[1-\frac{1}{4}\frac{2\cosh\varrho_0\sinh\varrho_0 + 15\varrho_0\tanh^2\varrho_0-3\varrho_0-15\tanh\varrho_0}{\cosh\varrho_1 \sinh\varrho_1 - 2\coth\varrho_1+ 3\varrho_1}\right]$\\
    %%%
    $\displaystyle \mathcal{A}=$
    & $\displaystyle \frac{1}{\mathcal{K}}
    \left[\sinh \varrho_0
        \left(1-\frac{1}{4}\frac{2\cosh\varrho_0\sinh\varrho_0 + 15\varrho_0 \tanh^2\varrho_0 - 3\varrho_0 -15\tanh\varrho_0}{\cosh\varrho_1\sinh\varrho_1-2\coth\varrho_1+3\varrho_1}\right)\right.$\\
     & $\displaystyle \left. 
    -{\rm sech} \varrho_0\left(1-\frac{1}{4}\frac{\cosh^2\varrho_0 + 6\varrho_0^2-15\varrho_0 \tanh\varrho_0}{\cosh\varrho_1\sinh\varrho_1-2\coth\varrho_1+3\varrho_1}\right)\right]$\\
    %%%
    $\displaystyle \mathcal{B}=$
    & $\displaystyle -\frac{1}{\mathcal{K}}\cosh\varrho_0 
    \left(1-\frac{1}{4}\frac{2\cosh\varrho_0\sinh\varrho_0 + 15\varrho_0 \tanh^2\varrho_0 - 3\varrho_0 -15\tanh\varrho_0}{\cosh\varrho_1\sinh\varrho_1-2\coth\varrho_1+3\varrho_1}\right)$\\
    %%%
    $\displaystyle \mathcal{C}=$
    & $\displaystyle \frac{1}{\mathcal{K}} {\rm sech} \varrho_0$\\
    %%%
    $\displaystyle \mathcal{D}=$
    & $\displaystyle -\frac{1}{\mathcal{K}}{\rm sech}\varrho_0\frac{1}{4}\frac{1}{\cosh\varrho_1\sinh\varrho_1-2\coth\varrho_1+3\varrho_1}$\\ \hline
    \end{tabular}
\end{table*}
%%%%%%%%%%%%%%%%%%%%%%%%%%%%%%%%%%%

%%%%%%%%%%%%%%%%%%%%%%%%%%%%%%%%%%%
\subsection{Stress fields around the intruder for $F_{\rm drag}\lesssim F_{\rm Y}$}
In this subsection, we focus on the stress fields around the intruder when the drag force $F_{\rm drag}$ is smaller than the threshold $F_{\rm Y}$, i.e., the intruder stops due to the balance between the drag force and those from the surrounding particles.

To characterize the stress, we adopt the similar method used in our previous paper \cite{Kubota21b}.
Once we adopt the theory of elasticity, the stress component is calculated from the simulation as
\begin{equation}
    \sigma_{\alpha\beta}
    \equiv \frac{1}{S}\sum_i
    \left(m_i v_{i,\alpha}v_{i,\beta}
    +\frac{1}{2}\sum_{j\neq i} r_{ij,\alpha} F_{ij,\beta}\right),
\end{equation}
with $\alpha, \beta=x, y, z$, where we have introduced the coarse graining method with the width $2d$ \cite{Zhang10}.
Now, it is convenient to consider the stress components in the elliptic coordinates $(\varrho, \vartheta)$ rather than the Cartesian coordinates $(x, y)$, either of which is converted from the other as
\begin{equation}
	x+iy = ic \cosh (\varrho + i \vartheta),\label{eq:def_elliptic_coordinates}
\end{equation}
with $c=\sqrt{R_1^2-R_2^2}$.
%%%%%%%%%%%%%%%%%%%%%%%%%%%%%%%%%%%
\begin{figure}[htbp]
    \centering
    \includegraphics[width=\linewidth]{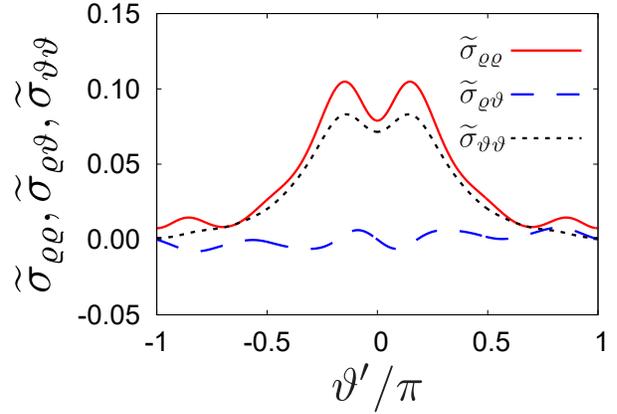}
    \caption{Stress profiles on the surface of the intruder against $\vartheta^\prime\equiv \vartheta+\pi/2$ for $\varphi=0.80$ and $F_{\rm drag}=1.58\times 10^{-2}k_n d$. 
    Solid, dotted, and dashed line represent $\widetilde{\sigma}_{\varrho\varrho}$, $\widetilde{\sigma}_{\vartheta\vartheta}$, and $\widetilde{\sigma}_{\varrho\vartheta}$, respectively.}
    \label{fig:stress_profile}
\end{figure}
%%%%%%%%%%%%%%%%%%%%%%%%%%%%%%%%%%%
Figure \ref{fig:stress_profile} shows typical profiles of the components of the stress on the surface of the intruder.

\begin{figure}[htbp]
    \centering
    \includegraphics[width=\linewidth]{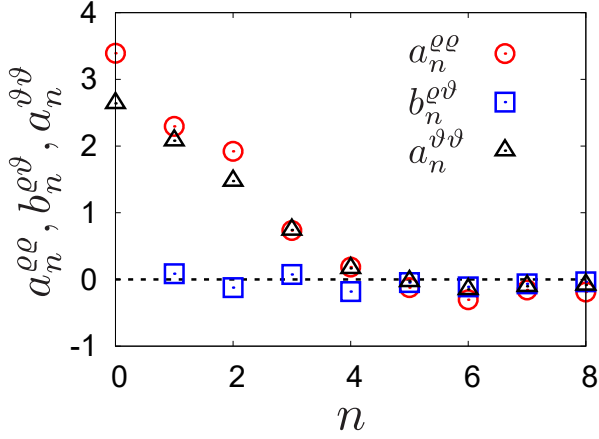}
    \caption{Fourier components of $\sigma_{\varrho\vartheta}$ for $\varphi=0.80$ and $F_{\rm drag}=1.58\times 10^{-2}k_n d$.
    Open circles, open squares, and open triangles represent $a_n^{\varrho\varrho}$, $b_n^{\varrho\vartheta}$, and $a_n^{\vartheta\vartheta}$, respectively.}
    \label{fig:stress_profile_fourier}
\end{figure}

To characterize the stress profiles, let us introduce the Fourier components of the stress on the surface of the intruder as
\begin{equation}
    \begin{Bmatrix}
        a_n^{(\varrho\vartheta)}\\
        b_n^{(\varrho\vartheta)}
    \end{Bmatrix}
    \equiv
    \frac{1}{\pi}\int_{-\pi}^\pi
    d\vartheta \frac{J(\varrho_0,\vartheta)^4}{c^2}
    \widetilde{\sigma}_{\varrho\vartheta}(\varrho_0,\vartheta)
    \begin{Bmatrix}
        \cos(n\vartheta)\\
        \sin(n\vartheta)
    \end{Bmatrix},
    \label{eq:Fourier_sigma}
\end{equation}
for $n=0,1,2,\cdots$, where we have introduced the normalized stress $\widetilde{\sigma}_{\varrho\vartheta} = \sigma_{\varrho\vartheta}/(F_{\rm drag}/\pi d)$. We also note that $b_0^{(\varrho\vartheta)}=0$ is always satisfied by its definition.
Figure \ref{fig:stress_profile_fourier} shows the behaviors of the Fourier components obtained from the simulation, where we have only shown $a_n^{(\varrho\varrho)}$, $b_n^{(\varrho\vartheta)}$, and $a_n^{(\vartheta\vartheta)}$ due to their symmetric properties.
The coefficients for $n\ge4$ become smaller than those for $n\le 3$, which suggests that it is sufficient to consider terms up to the third-order even when we consider the stress around the intruder.

%%%%%%%%%%%%%%%%%%%%%%%%%%%%%%%%%%%
%%%%%%%%%%%%%%%%%%%%%%%%%%%%%%%%%%%

Once we adopt this assumption, the expressions of the stress components in the elliptic coordinates are listed in Table \ref{fig:sigma_elliptic}, where we have introduced $\vartheta^\prime\equiv \vartheta+\pi/2$ (see Appendix \ref{sec:elliptic_stressfield} for details).
%%%%%%%%%%%%%%%%%%%%%%%%%%%%%%%%%%%
\begin{table*}[htbp]
    \caption{The expressions of the stress components in the elliptic coordinates: $\widetilde{\sigma}_{\varrho\varrho}$, $\widetilde{\sigma}_{\varrho\vartheta}$, and $\widetilde{\sigma}_{\vartheta\vartheta}$.}
    \label{fig:sigma_elliptic}
    \centering
    \begin{tabular}{ll}\hline
    $\displaystyle \frac{J^4}{c^2}\widetilde{\sigma}_{\varrho\varrho}=$
    & $\displaystyle \frac{e^{2\varrho} - e^{-2\varrho}}{4}E 
        + \frac{2 + e^{-4\varrho}}{2}F
        + \frac{3}{2}e^{-2\varrho}A_2
        + \frac{3}{2}e^{-4\varrho}B_2$\\
     & $\displaystyle +\left[
        -\frac{e^{3\varrho} + e^{-3\varrho} - e^{\varrho} - e^{-\varrho}}{8}G
        + \frac{2\left(e^{3\varrho} + e^{-3\varrho}\right) + 5\left(e^{\varrho} + e^{-\varrho}\right)}{8}H
        + \frac{e^{\varrho} + e^{-\varrho}}{2}C_1
    \right.$\\
     & $\displaystyle \hspace{1.5em}\left.
        - 3e^{-3\varrho}C_3
        - \frac{4e^{-\varrho} - e^{-3\varrho} + e^{-5\varrho}}{2}D_1
        - 3e^{-5\varrho}D_3
    \right]\cos(\vartheta^\prime)$\\
     & $\displaystyle +\left[
        \left(e^{2\varrho}+e^{-2\varrho}\right)F
        + \frac{3+e^{-4\varrho}}{2}A_2 
        + 2e^{-2\varrho}B_2
    \right]\cos(2\vartheta^\prime)$\\
     & $\displaystyle +\left[
        \frac{e^{\varrho} + e^{-\varrho}}{8}H
        - \frac{6e^{-\varrho}+3e^{-5\varrho}}{2}C_3
        - \frac{5e^{\varrho}+4e^{-3\varrho}}{2}D_1
        - \frac{7e^{-3\varrho}+2e^{-7\varrho}}{2}D_3
    \right]\cos(3\vartheta^\prime)$\\ \hline
    %%%
    %%%
    %%%
    $\displaystyle \frac{J^4}{c^2}\widetilde{\sigma}_{\varrho\vartheta} =$ & 
    $\displaystyle \left[
        \frac{3e^{\varrho} - 3e^{-\varrho} - e^{3\varrho} + e^{-3\varrho}}{8}G
        + \frac{3e^{\varrho}-3e^{-\varrho}}{8}H 
        + \frac{e^{\varrho} - e^{-\varrho}}{2}C_1
        - 3e^{-3\varrho}C_3 
    \right.$\\
     & $\displaystyle \hspace{0.5em}\left.
        + \frac{-3e^{-\varrho}+e^{-5\varrho}}{2}D_1 
        - 5e^{-5\varrho}D_3
    \right]\sin(\vartheta^\prime)$\\
     & $\displaystyle+\left[
        \frac{1}{2}E 
        + \frac{e^{2\varrho}+e^{-2\varrho}}{2}F 
        + \frac{3+e^{-4\varrho}}{2}A_2 
        + \frac{5e^{-2\varrho} + 3e^{-6\varrho}}{2}B_2
     \right]\sin(2\vartheta^\prime)$\\
     & $\displaystyle +\left[
        -\frac{e^{\varrho} - e^{-\varrho}}{8}H 
        - \frac{6e^{-\varrho} + 3e^{-5\varrho}}{2}C_3 
        - \frac{3}{2}e^{\varrho}D_1 
        - \frac{9e^{-3\varrho} + 6e^{-7\varrho}}{2}D_3 
    \right]\sin(3\vartheta^\prime)$\\ \hline
    %%%
    %%%
    %%%
    $\displaystyle \frac{J^4}{c^2}\widetilde{\sigma}_{\vartheta\vartheta} =$ &
    $\displaystyle - \frac{e^{2\varrho} - e^{-2\varrho}}{4}E
        + \frac{1}{2}\left(2 + e^{-4\varrho}\right)F
        - \frac{3}{2}e^{-2\varrho}A_2
        - \frac{9}{2}e^{-4\varrho}B_2$\\
     & $\displaystyle +\left[
        -\frac{5\left(e^{\varrho} + e^{-\varrho}\right) + e^{3\varrho} + e^{-3\varrho}}{8}G
        - \frac{e^{\varrho} + e^{-\varrho}}{8}H
        - \frac{e^{\varrho} + e^{-\varrho}}{2}C_1
        + 3e^{-3\varrho}C_3
    \right.$\\
     & $\displaystyle \hspace{1.5em}\left.
        - \frac{4e^{-\varrho} + 5e^{-3\varrho} + 3e^{-5\varrho}}{2}D_1
        + 7e^{-5\varrho}D_3
    \right]\cos(\vartheta^\prime)$\\
     & $\displaystyle +\left[
        2e^{-2\varrho}F 
        - \frac{3 + 2e^{-4\varrho}}{2}A_2 
        - \left(5e^{-2\varrho} + 3e^{-6\varrho}\right)B_2
    \right]\cos(2\vartheta^\prime)$\\
     & $\displaystyle +\left[
        - \frac{e^{\varrho} + e^{-\varrho}}{4}G
        + \frac{e^{\varrho} + e^{-\varrho}}{8}H
        + \frac{6e^{-\varrho} + 3e^{-5\varrho}}{2}C_3\right.$\\
     & $\displaystyle \hspace{1.5em}\left.
        + \frac{e^{\varrho} - 4e^{-3\varrho}}{2}D_1
        + \frac{15e^{-3\varrho} + 10e^{-7\varrho}}{2}D_3
    \right]\cos(3\vartheta^\prime)$ \\ \hline
    \end{tabular}
\end{table*}
%%%%%%%%%%%%%%%%%%%%%%%%%%%%%%%%%%%
In Table \ref{fig:sigma_elliptic}, we have ten unknown coefficients; $E$, $F$, $G$, $H$, $A_2$, $B_2$, $C_1$, $C_3$, $D_1$, and $D_3$.
Now, let determine these coefficients from the information of them on the surface of the intruder \cite{Kubota21b}.
From Eq.\ \eqref{eq:Fourier_sigma} and Table \ref{fig:sigma_elliptic}, the following equation is satisfied on the surface of the intruder:
\begin{equation}
    \mathcal{M}\bm{X}=\widetilde{\bm{\Sigma}}_{\rm surf},
    \label{eq:CX_Sigma}
\end{equation}
where the vectors $\bm{X}$ and $\widetilde{\bm{\Sigma}}_{\rm surf}$ are defined by
\begin{align}
    &\bm{X}
    \equiv (E, F, G, H, A_2, B_2, C_1, C_3, D_1, D_3)^T,\\
    %%%
    &\widetilde{\bm{\Sigma}}_{\rm surf}
    \equiv (a_0^{\varrho\varrho},a_1^{\varrho\varrho},a_2^{\varrho\varrho},a_3^{\varrho\varrho},b_1^{\varrho\vartheta},b_2^{\varrho\vartheta},b_3^{\varrho\vartheta},\nonumber\\
    &\hspace{4em} a_0^{\vartheta\vartheta},a_1^{\vartheta\vartheta},a_3^{\vartheta\vartheta})^T,
\end{align}
respectively, and the each component of the matrix $\mathcal{M}$ (the size of $\mathcal{M}$ is $10\times10$) is listed in Table \ref{fig:components_M}.

%%%%%%%%%%%%%%%%%%%%%%%%%%%%%%%%%%%
\begin{table*}[htbp]
    \caption{Each component of $\mathcal{M}$.}
    \label{fig:components_M}
    \centering
    \begin{tabular}{lll}\hline
    $\displaystyle \mathcal{M}_{1,1} = \frac{e^{2\varrho} - e^{-2\varrho}}{4}$
     & $\displaystyle \mathcal{M}_{1,2} = \frac{2 + e^{-4\varrho}}{2}$
     & $\displaystyle \mathcal{M}_{1,5} = \frac{3}{2}e^{-2\varrho}$\\
    $\displaystyle \mathcal{M}_{1,6} = \frac{3}{2}e^{-4\varrho}$
    %%%%%
     & $\displaystyle \mathcal{M}_{2,3} = \frac{e^{\varrho} + e^{-\varrho} - e^{3\varrho} - e^{-3\varrho}}{8}$
     & $\displaystyle \mathcal{M}_{2,4} = \frac{5e^{\varrho} + 5e^{-\varrho} + 2e^{3\varrho} + 2e^{-3\varrho}}{8}$\\
    $\displaystyle \mathcal{M}_{2,7} = \frac{e^{\varrho} + e^{-\varrho}}{2}$
     & $\displaystyle \mathcal{M}_{2,8} = -3e^{-3\varrho}$
     & $\displaystyle \mathcal{M}_{2,9} = -\frac{4e^{-\varrho} - e^{-3\varrho} + e^{-5\varrho}}{2}$\\
    $\displaystyle \mathcal{M}_{2,10} = \displaystyle -3e^{-5\varrho}$
    %%%%%
     & $\displaystyle \mathcal{M}_{3,2} = e^{2\varrho} + e^{-2\varrho}$
     & $\displaystyle \mathcal{M}_{3,5} = \frac{3 + e^{-4\varrho}}{2}$\\
    $\displaystyle \mathcal{M}_{3,6} = 2e^{-2\varrho}$
    %%%%%
     & $\displaystyle \mathcal{M}_{4,4} = \frac{e^{\varrho} + e^{-\varrho}}{8}$
     & $\displaystyle \mathcal{M}_{4,8} = -\frac{6e^{-\varrho} + 3e^{-5\varrho}}{2}$\\
    $\displaystyle \mathcal{M}_{4,9} = -\frac{5e^{\varrho} + 4e^{-3\varrho}}{2}$
     & $\displaystyle \mathcal{M}_{4,10} = -\frac{7e^{-3\varrho} + 2e^{-7\varrho}}{2}$
    %%%%%
     & $\displaystyle \mathcal{M}_{5,3} = \frac{3e^{\varrho} - 3e^{-\varrho} - e^{3\varrho} + e^{-3\varrho}}{8}$\\
    $\displaystyle \mathcal{M}_{5,4} = \frac{3e^{\varrho} - 3e^{-\varrho}}{8}$
     & $\displaystyle \mathcal{M}_{5,7} = \frac{e^{\varrho} - e^{-\varrho}}{2}$
     & $\displaystyle \mathcal{M}_{5,8} = -3e^{-3\varrho}$\\
    $\displaystyle \mathcal{M}_{5,9} = -\frac{3e^{-\varrho} - e^{-5\varrho}}{2}$
     & $\displaystyle \mathcal{M}_{5,10} = -5e^{-5\varrho}$
    %%%%%
     & $\displaystyle \mathcal{M}_{6,1} = -\frac{1}{2}$\\
    $\displaystyle \mathcal{M}_{6,2} = \frac{e^{2\varrho} + e^{-2\varrho}}{8}$
     & $\displaystyle \mathcal{M}_{6,5} = \frac{3 + e^{-4\varrho}}{2}$
     & $\displaystyle \mathcal{M}_{6,6} = \frac{5e^{-2\varrho} + 3e^{-6\varrho}}{2}$\\
    %%%%%
    $\displaystyle \mathcal{M}_{7,4}= -\frac{e^{\varrho} - e^{-\varrho}}{8}$
     & $\displaystyle \mathcal{M}_{7,8}= -\frac{6e^{-\varrho} + 3e^{-5\varrho}}{2}$
     & $\displaystyle \mathcal{M}_{7,9}= -\frac{3}{2}e^{\varrho}$\\
    $\displaystyle \mathcal{M}_{7,10}= -\frac{9e^{-3\varrho} + 6e^{-7\varrho}}{2}$
    %%%%%%
     & $\displaystyle \mathcal{M}_{8,1}= -\frac{e^{2\varrho} - e^{-2\varrho}}{4}$
     & $\displaystyle \mathcal{M}_{8,2}= -\frac{2 + e^{-4\varrho}}{2}$\\
    $\displaystyle \mathcal{M}_{8,5}= -\frac{3}{2}e^{-2\varrho}$
     & $\displaystyle \mathcal{M}_{8,6}= -\frac{9}{2}e^{-4\varrho}$
    %%%%%
     & $\displaystyle \mathcal{M}_{9,3}= -\frac{5e^{\varrho} + 5e^{-\varrho} + e^{3\varrho} + e^{-3\varrho}}{8}$\\
    $\displaystyle \mathcal{M}_{9,4}= -\frac{e^{\varrho} + e^{-\varrho}}{8}$
     & $\displaystyle \mathcal{M}_{9,7}= -\frac{e^{\varrho} + e^{-\varrho}}{2}$
     & $\displaystyle \mathcal{M}_{9,8}= 3e^{-3\varrho}$\\
    $\displaystyle \mathcal{M}_{9,9}= -\frac{4e^{-\varrho} + 5e^{-3\varrho} + 3e^{-5\varrho}}{2}$
     & $\displaystyle \mathcal{M}_{9,10}= 7e^{-5\varrho}$
    %%%%%
     & $\displaystyle \mathcal{M}_{10,3}= -\frac{e^{\varrho} + e^{-\varrho}}{4}$\\
    $\displaystyle \mathcal{M}_{10,4}= \frac{e^{\varrho} + e^{-\varrho}}{8}$
     & $\displaystyle \mathcal{M}_{10,8}= \frac{6e^{-\varrho} + 3e^{-5\varrho}}{2}$
     & $\displaystyle \mathcal{M}_{10,9}= \frac{e^{\varrho} - 4e^{-3\varrho}}{2}$\\
    $\displaystyle \mathcal{M}_{10,10}= \frac{15e^{-3\varrho} + 10e^{-7\varrho}}{2}$\\ \hline
    \end{tabular}
\end{table*}
%%%%%%%%%%%%%%%%%%%%%%%%%%%%%%%%%%%

\begin{figure}[htbp]
    \centering
    \includegraphics[width=\linewidth]{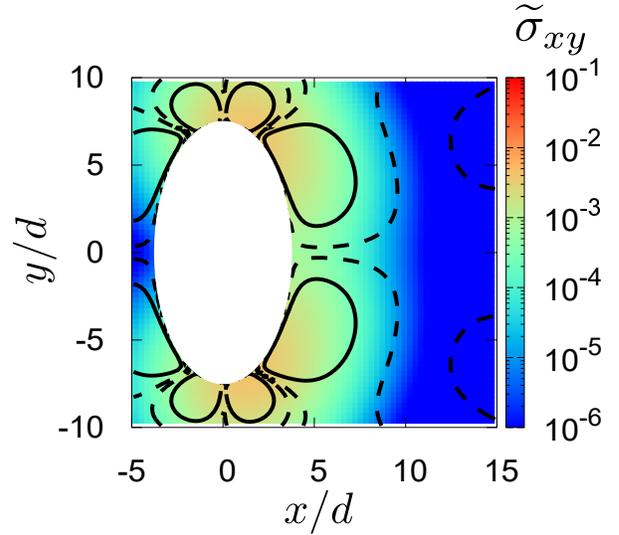}
    \caption{Absolute value of the shear stress $|\sigma_{xy}|$ around the intruder (white ellipse) for $\varphi=0.80$ and $F_{\rm drag}=1.58\times 10^{-2}k_n d$.
    The solid and dashed lines represent the contours for $5.0\times10^{-2}$ and $1.0\times10^{-2}$, respectively.}
    \label{fig:sigma}
\end{figure}

Now, we can numerically solve Eq.\ \eqref{eq:CX_Sigma} as
\begin{equation}
    \bm{X}=\mathcal{M}^{-1}\widetilde{\bm{\Sigma}}_{\rm surf}.
\end{equation}
Using this solution, we can evaluate the stress fields.
Figure \ref{fig:sigma} shows the comparison of $\sigma_{xy}$ obtained from the simulation and that from the expressions listed in Table \ref{fig:sigma_elliptic} with Eqs.\ \eqref{eq:sigma_xy_alphabeta}.

%%%%%%%%%%%%%%%%%%%%%%%%%%%%%%%%%%%
%%%%%%%%%%%%%%%%%%%%%%%%%%%%%%%%%%%

%%%%%%%%%%%%%%%%%%%%%%%%%%%%%%%%%%%
\section{Discussion and conclusion}\label{sec:conclusion}
In this paper, we have numerically investigated the drag of an elliptic intruder in a two-dimensional granular environment.
First, the drag law is reproduced by the sum of the yield force and the dynamic part, where the former is determined from the Coulombic friction between the particles and the bottom plate.
On the other hand, the latter (dynamic part) is understood from a collision model, where the surrounding particles collide with the intruder at a steady speed, which is consistent with the previous studies.
Next, the streamlines are better reproduced by the perfect fluid than by the viscous flow in from of the intruder, while neither captures the flow in the back of the intruder.
Third, we consider the drag below the yield force.
The stress fields acting around the intruder can be reproduced from the drag force acting on the surface of the intruder by using the two-dimensional theory of elasticity.

The following remains the future perspective:
In this study, the intruder was only exposed to translational force. 
The effect of the force perpendicular to the drag force might be important.
Next, we have only considered a simple elliptic shape as the intruder. 
The asymmetric shape might affect the drag law, which might be important in industrial applications.
Last, we have focused on two-dimensional systems.
The three-dimensional system might exhibit different behavior such as vortex motions.

%%%%%%%%%%%%%%%%%%%%%%%%%%%%%%%%%%%
\section*{Acknowledgments}
This work is financially supported by the Grant-in-Aid of MEXT for Scientific Research (Grant Nos.\ \href{https://kaken.nii.ac.jp/en/grant/KAKENHI-PROJECT-20K14428}{JP20K14428} and \href{https://kaken.nii.ac.jp/en/grant/KAKENHI-PROJECT-21H01006/}{JP21H01006}).

%%%%%%%%%%%%%%%%%%%%%%%%%%%%%%%%%%%
\appendix
%%%%%%%%%%%%%%%%%%%%%%%%%%%%%%%%%%%
\section{Detection of collision between intruder and surrounding particles}\label{sec:coll_detection}
In this Appendix, we explain the treatment of a collision between the intruder and a surrounding particle.
We note that the main procedure in this Appendix follows that reported in Ref.\ \cite{Dziugys01}.
We put $\bm{r}_0=(x_0, y_0)$, $\bm{r}_i=(x_i, y_i)$, and $\theta_0$ as the positions of the intruder and the particle, respectively, and the angle of the major axis of the intruder.

First, we move the intruder to the origin and rotate with the angle $-\theta_0$.
Correspondingly, the position of the particle becomes
\begin{equation}
\begin{cases}
  x_i^\prime = (x_i-x_0)\cos\theta_0 + (y_i-y_0)\sin\theta_0,\\
  y_i^\prime = -(x_i-x_0)\sin\theta_0 + (y_i-y_0)\cos\theta_0.
\end{cases}
\end{equation}
Then, it is obvious that a collision occurs when the following condition satisfies:
\begin{equation}
  \frac{x_i^{\prime2}}{(R_1+d_i/2)^2} + \frac{y_i^{\prime2}}{(R_2+d_i/2)^2}\le 1.
\end{equation}
Under this condition, we seek for a point $(x, y)$ on the surface of the elliptic intruder, which is closest to the center of the particle $(x_i^\prime, y_i^\prime)$.
It should be noted that the former point satisfies 
\begin{equation}
  \frac{x^2}{R_1^2} + \frac{y^2}{R_2^2} = 1.
  \label{eq:eq_ellipse}
\end{equation}
Let us introduce the distance between these two points as
\begin{align}
    r_{\rm min}^2 
    &\equiv (x-x_i)^2 + (y-y_i)^2\nonumber\\
    &= (x-x_i)^2 + \left(\pm \frac{R_2}{R_1}\sqrt{R_1^2-x^2} - y_i\right)^2\nonumber\\
    &= \left(1-\frac{R_2^2}{R_1^2}\right)x^2 - 2x_i x \mp 2\frac{R_2}{R_1}y_i\sqrt{R_1^2-x^2}\nonumber\\
    &\hspace{1em}+x_i^2 + y_i^2 + R_2^2.
\end{align}
where we have used Eq.\ \eqref{eq:eq_ellipse} in the second line.
For a closest point, this distance satisfies $\partial r_{\rm min}^2/\partial x =0$, that is,
\begin{equation}
    2\left[\left(1-\frac{R_2^2}{R_1^2}\right)x - x_i
    \pm\frac{R_2}{R_1}\frac{y_ix}{\sqrt{R_1^2-x^2}}\right]
    =0.
\end{equation}
This yields a following quartic equation:
\begin{align}
    &\left(1-\frac{R_2^2}{R_1^2}\right)^2x^4
    -2\left(1-\frac{R_2^2}{R_1^2}\right)x_i x^3\nonumber\\
    &+\left[x_i^2 + \frac{R_2^2}{R_1^2}y_i^2 - R_1^2\left(1-\frac{R_2^2}{R_1^2}\right)^2\right]x^2\nonumber\\
    &+2R_1^2 \left(1-\frac{R_2^2}{R_1^2}\right)x_ix -R_1^2x_i^2 =0.
    \label{eq:quintic_eq}
\end{align}
Solving this equation, we can determine the point at which the distance becomes shortest.
It should be noted that we always obtain only one solution which satisfies $x\le |x_i|$.
Practically, it is convenient to introduce 
\begin{equation}
  X \equiv \frac{x}{R_1},\quad
  X_i \equiv \frac{R_1}{R_1^2-R_2^2}x_i,\quad
  Y_i \equiv \frac{R_2}{R_1^2-R_2^2}y_i.
\end{equation}
Then, the quartic equation \eqref{eq:quintic_eq} is rewritten as
\begin{equation}
  X^4 - 2X_i X^3 + (X_i^2+Y_i^2-1)X^2+2X_i X - X_i^2=0.
  \label{eq:quintic_eq_2}
\end{equation}
Once, we obtain a solution $(x_{\rm c}^\prime, y_{\rm c}^\prime)$ of Eq.\ \eqref{eq:quintic_eq_2}, the actual position $(x_{\rm c}, y_{\rm c})$ is given by
\begin{equation}
\begin{cases}
  x_{\rm c}=x_0 + x_{\rm c}^\prime \cos\theta_0 - y_{\rm c}^\prime \sin\theta_0,\\
  y_{\rm c}=y_0 + x_{\rm c}^\prime \sin\theta_0 + y_{\rm c}^\prime \cos\theta_0.
\end{cases}
\end{equation}
Then, the normal vector from the surface of the intruder to the center of the particle $\bm{n}$ becomes
\begin{equation}
  \bm{n}=(n_x, n_y)^T,
\end{equation}
with
\begin{subequations}
\begin{align}
    n_x &\equiv \frac{x_i-x_{\rm c}}{\sqrt{(x_i-x_{\rm c})^2+(y_i-y_{\rm c})^2}},\\
    n_y &\equiv \frac{y_i-y_{\rm c}}{\sqrt{(x_i-x_{\rm c})^2+(y_i-y_{\rm c})^2}}.
\end{align}
\end{subequations}
%%%%%%%%%%%%%%%%%%%%%%%%%%%%%%%%%%%

%%%%%%%%%%%%%%%%%%%%%%%%%%%%%%%%%%%
\section{Streamlines of two-dimensional viscous fluid}\label{sec:viscous_elliptic}
In this Appendix, we summarize the expression of the stream function of the two-dimensional viscous fluid based on Ref.\ \cite{Raynor02}.
Let us introduce $(\varrho, \vartheta)$ as
\begin{equation}
\begin{cases}
  \varrho = {\rm Re}\left\{\log \left[i\left(z \pm \sqrt{z^2+1}\right)\right]\right\},\\
  \vartheta = {\rm Im}\left\{\log \left[i\left(z \pm \sqrt{z^2+1}\right)\right]\right\},
\end{cases}
  \label{eq:varrho_vartheta}
\end{equation}
where $z=x+iy$.
The sign should be selected considering the Riemann surface.
Here, the surface of the intruder is characterized by $\varrho\equiv\varrho_0$ with
\begin{equation}
  \varrho_0 = \frac{1}{2}\log \frac{R_1+R_2}{R_1-R_2}.
\end{equation}

The stream function $\Psi$ of the viscous fluid is known to satisfy the biharmonic equation
\begin{equation}
  \Delta \Delta \Psi=0,
\end{equation}
where the Laplacian operator $\Delta\equiv \partial^2/\partial x^2+\partial^2/\partial y^2$ is defined in the $(x,y)$ plane.
We solve this biharmonic equation using the following boundary conditions:
\begin{equation}
\begin{cases}
  v_\varrho = v_\vartheta = \omega = 0 & (\varrho=\varrho_0),\\
  \Psi = Uy & (\varrho=\varrho_1).
\end{cases}
\end{equation}
It should be noted that the outer boundary $\varrho_1$ should be sufficiently far from the intruder, i.e., $\varrho_1\gg\varrho_0$.
After the same procedure as that in Ref.\ \cite{Raynor02}, we obtain the stream function as Eq.\ \eqref{eq:Psi_viscous}, where we have introduced the coefficients $\mathcal{K}$, $\mathcal{A}$, $\mathcal{B}$, $\mathcal{C}$, and $\mathcal{D}$ as listed in Table \ref{fig:KABCD}.

%%%%%%%%%%%%%%%%%%%%%%%%%%%%%%%%%%%
\section{Stress fields in the elliptic coordinates}\label{sec:elliptic_stressfield}
In this Appendix, we briefly explain the procedure to derive the stress around an intruder from the theory of elasticity.
First, let us consider ($\varrho$, $\vartheta$) space which is connected from the Cartesian coordinates as
\begin{equation}
	z = ic \cosh \zeta
	\label{eq:z_zeta}
\end{equation}
with $z=x+iy$ and $\zeta=\varrho+i\vartheta$, or equivalently,
\begin{subequations}\label{eq:xy_alphabeta}
\begin{align}
	x &= -c \sinh \varrho \sin \vartheta,\\
	y &= c \cosh \varrho \cos \vartheta,
\end{align}
\end{subequations}
where we have introduced $z=x+iy$ and $\zeta=\varrho+i\vartheta$ with a constant $c$.
Also, by changing the form of Eq.\ \eqref{eq:xy_alphabeta}, we can obtain 
\begin{subequations}\label{eq:xy_alphabeta_2}
\begin{align}
  \frac{x^2}{c^2 \sinh^2\varrho} + \frac{y^2}{c^2 \cosh^2\varrho} = 1,\\
  -\frac{x^2}{c^2 \sin^2\vartheta} + \frac{y^2}{c^2 \cos^2\vartheta} = 1.
\end{align}
\end{subequations}
These equations correspond to an elliptic and a parabolic when we fix $\varrho$ and $\vartheta$ as constants, respectively.

Now, we define the expansion ratio $J$ as follows:
\begin{equation}
	J = \sqrt{x_\varrho^2 + x_\vartheta^2},
\end{equation}
where $x_\varrho=(\partial x/\partial \varrho)$ and $x_\vartheta=(\partial x/\partial \vartheta)$ are, respectively, given by
\begin{align}
	x_\varrho &= -c \cosh \varrho \sin \vartheta,\\
	x_\vartheta &= -c \sinh \varrho \cos \vartheta.
\end{align}
From these equations, the ratio $J$ can be written as
\begin{align}
	J^2 = x_\varrho^2 + x_\vartheta^2
	&= \frac{1}{2}c^2 \left(\cosh{2\varrho} - \cos{2\vartheta}\right).
\end{align}
Next, let us calculate the angle $\gamma$ between the $\varrho$ and the $x$ directions at the point of ($x$, $y$). 
This can be easily obtained by considering the infinitesimal change of $\varrho$ under the condition $\vartheta={\rm const.}$, which yields the following relations:
\begin{align}
	&(\cos\gamma, \sin\gamma)
	= \left(\frac{x_\varrho}{\sqrt{x_\varrho^2+y_\varrho^2}}, 
      \frac{y_\varrho}{\sqrt{x_\varrho^2+y_\varrho^2}}\right)\nonumber\\
	&= \left(-\frac{\sqrt{2}\cosh\varrho\sin\vartheta}{\sqrt{\cosh{2\varrho} - \cos{2\vartheta}}}, 
      \frac{\sqrt{2}\sinh\varrho\cos\vartheta}{\sqrt{\cosh{2\varrho} - \cos{2\vartheta}}}\right).
	\label{eq:bipolar_cos_sin}
\end{align}
From these, once we can obtain the stress components in the elliptic coordinates, they are converted into those in the Cartesian coordinates as
\begin{subequations}\label{eq:sigma_xy_alphabeta}
\begin{align}
	\sigma_{xx} 
	&= \sigma_{\varrho\varrho}\cos^2\gamma + \sigma_{\vartheta\vartheta}\sin^2\gamma - 2\sigma_{\varrho\vartheta}\sin\gamma\cos\gamma,\\
	\sigma_{yy}
	&= \sigma_{\varrho\varrho}\sin^2\gamma + \sigma_{\vartheta\vartheta}\cos^2\gamma + 2\sigma_{\varrho\vartheta}\sin\gamma\cos\gamma,\\
	\sigma_{xy}
	&= (\sigma_{\varrho\varrho}-\sigma_{\vartheta\vartheta})\sin\gamma\cos\gamma + \sigma_{\varrho\vartheta}(\cos^2\gamma-\sin^2\gamma),
\end{align}
\end{subequations}
respectively.

It is well known that each component of the stress in the static problem can be expressed by the Airy stress function $\chi$ \cite{Timoshenko,Daniels17}.
Now, let us consider the biharmonic equation in the Cartesian coordinates:
\begin{equation}
	\Delta \Delta \chi(x,y)=0.
\end{equation}
Once we introduce the elliptic coordinates, the general solution of this equations becomes
\begin{align}
  &\chi(\varrho,\vartheta)\nonumber\\
  &= E\varrho + F e^{-2\varrho} + F \cos(2\vartheta)+ G \varrho\sinh\varrho\sin\vartheta \nonumber\\
    &\hspace{1em} + H\vartheta\cosh\varrho\cos\vartheta +C_1^\prime e^{\varrho}\sin\vartheta\nonumber\\
    &\hspace{1em} + \sum_{n:{\rm even}} \left(A_n+B_n e^{-2\varrho} + B_{n-2}e^{2\varrho}\right)e^{-n\varrho}\cos(n\vartheta)\nonumber\\
    &\hspace{1em} + \sum_{n: {\rm odd}} \left(C_n+D_n e^{-2\varrho} + D_{n-2}e^{2\varrho}\right)e^{-n\varrho}\sin(n\vartheta),
\end{align}
where we have dropped some terms which diverge at $\varrho\to\infty$.
Now, we assume that it is sufficient to consider terms up to the third order of $n$.
In addition, let us focus on the region in front of the intruder.
This means that the stress should converge to zero for $\varrho\to\infty$.
After some calculations, it is sufficient to consider the following form of the Airy functions:
\begin{align}
    &\chi(\varrho,\vartheta)\nonumber\\
    &= E\varrho + F e^{-2\varrho} + G \varrho\sinh\varrho\sin\vartheta 
    + H\vartheta\cosh\varrho\cos\vartheta\nonumber\\
    &\hspace{1em} +(F + A_2e^{-2\varrho}+B_2e^{-4\varrho})\cos(2\vartheta)\nonumber\\
    &\hspace{1em} + (C_1\cosh\varrho + D_1 e^{-3\varrho})\sin\vartheta \nonumber\\
    &\hspace{1em} + (D_1e^{-\varrho} + C_3e^{-3\varrho} + D_3e^{-5\varrho}) \sin(3\vartheta),
    \label{eq:Psi_assumption}
\end{align}
where asymmetric parts are also dropped with respect to $y\leftrightarrow -y$.
This function has ten unknown coefficients; $E$, $F$, $G$, $H$, $A_2$, $B_2$, $C_1$, $C_3$, $D_1$, and $D_3$, which will be determined by the information of the stress profile on the surface of the intruder.
From Eq.\ \eqref{eq:Psi_assumption}, each component of the stress is given by Table \ref{fig:sigma_elliptic} with $\vartheta^\prime=\vartheta+\pi/2$.

%%%%%%%%%%%%%%%%%%%%%%%%%%%%%%%%%%%
\section*{Nomenclature}
\noindent
\begin{tabular}{p{27pt}p{17em}}
$a$ & coefficient of elliptic coordinates (N/m)\\
$a_n^{\alpha\beta}$ & Fourier components of stress ($\sigma_{\alpha\beta}$) in elliptic coordinates (-)\\
$\mathcal{A}$ & component of stream function of the viscous fluid (m$^2$/s)\\
$A_2$ & component of Airy stress function (-)\\
$b_n^{\alpha\beta}$ & Fourier components of stress ($\sigma_{\alpha\beta}$) in elliptic coordinates (-)\\
$\mathcal{B}$ & component of stream function of the viscous fluid (m$^2$/s)\\
$B_2$ & component of Airy stress function (-)\\
$c$ & coefficient of polar coordinates (m)\\
$\mathcal{C}$ & component of stream function of the viscous fluid (m$^2$/s)\\
$C_1$ & component of Airy stress function (-)\\
$C_3$ & component of Airy stress function (-)\\
$d_i$ & diameter of $i$-th particle (m) \\
$d_{ij}$ & sum of diameters of $i$-th and $j$-th particles (m)\\
$\mathcal{D}$ & component of stream function of the viscous fluid (m$^2$/s)\\
$D_1$ & component of Airy stress function (-)\\
$D_3$ & component of Airy stress function (-)\\
$e$ & repulsion coefficient (-)\\
$\hat{\bm{e}}_x$ & unit vector along the $x$-axis (-)\\
$E$ & component of Airy stress function (-)\\
$F$ & component of Airy stress function (-)\\
$F_\mathrm{drag}$ & drag force acting on the intruder (N)\\
$F_{ij}$ & interaction between $i$-th and $j$-th particles (N)\\
$F_{\rm Y}$ & yield force (N)\\
$g$ & gravitational acceleration (m/s$^2$)\\
$G$ & component of Airy stress function (-)\\
$H$ & component of Airy stress function (-)\\
$I_0$ & moment of inertia of intruder (kg$\cdot$m$^2$)\\
$I_i$ & moment of inertia of $i$-th particle (kg$\cdot$m$^2$)\\
$J$ & expansion ratio (m$^{1/2}$)\\
$k_n$ & spring constant in the normal direction (N/m)
\end{tabular}

\begin{tabular}{p{27pt}p{17em}}$k_t$ & spring constant in the tangential direction (N/m)\\
$\mathcal{K}$ & component of stream function of the viscous fluid (m$^2$/s)\\
$L_y$ & system length in $y$-direction (m)\\
$m_i$ & mass of $i$-th particle (kg) \\
$m_\mathrm{r}$ & reduced mass (kg)\\
$M$ & mass of the intruder (kg) \\
$\mathcal{M}$ & matrix of the coefficients (-)\\
$\bm n$ & unit vector for the normal direction between contacting particles (-)\\
$N$ & number of particles (-)\\
$p$ & impulse at collision (N$\cdot$s)\\
$\bm{r}_i$ & position vector of $i$-th particle (m)\\
$r_{ij}$ & relative displacement between $i$-th and $j$-th particles (m)\\
$\hat{\bm r}_{ij}$ & unit vector of relative displacement between $i$-th and $j$-th particles (-)\\
$R$ & radius of the cylinder (m)\\
$R_1$ & half length along to the major axis of the intruder (m)\\
$R_2$ & half length along to the minor axis of the intruder (m)\\
$S$ & area to be coarse grained (m$^2$)\\
$t$ & time (s)\\
$\bm t$ & unit vector for the tangential direction between contacting particles (-)\\
$T_i$ & torque acting on $i$-th particle (N$\cdot$m)\\
$U$ & flow velocity (m/s)\\
$\bm{v}_i$ & velocity vector of $i$-th particle (m/s)\\
$\bm {v}_{ij}$ & relative velocity vector between $i$-th and $j$-th particles (m/s)\\
$\bm {v}_{ij}$ & tangential component of the relative velocity vector (m/s)\\
$V$ & steady speed of the intruder (m/s)\\
$\mathcal{V}_\mathrm{coll}$ & (two-dimensinal) volume of the collision cylinder (m$^2$)\\
$W(z)$ & complex velocity function (m/s)\\
$\bm X$ & vectors of the unknown coefficients (-)\\
$z$ & value of elliptic coordinates (-)
\end{tabular}

\begin{tabular}{p{27pt}p{17em}}
$\alpha$ & coefficient of the dynamic part of the drag force (kg/m)\\
$\beta$ & angle between $x$-axis and the normal at a point of the intruder surface (-)\\
$\gamma$ & angle between the $\rho$ and $x$ directions (-)\\
$\delta$ & overlap between contacting particles (m)\\
$\delta_{i,j}$ & Kronecker delta (-)\\
$\epsilon$ &eccentricity (-)\\
$\zeta$ & polar coordinates (-)\\
$\eta_n$ & damping coefficient in the normal direction (N$\cdot$s/m)\\
$\eta_t$ & damping coefficient in the tangential direction (N$\cdot$s/m)\\
$\theta$ & angle of the intruder (-)\\
$\vartheta$ & component of the elliptic coordinate (-)\\
$\theta_i$ & angle of $i$-th particle (-)\\
$\Theta$ & step function (-)\\
$\mu$ & tangential friction coefficient between contacting particles (-)\\
$\mu_\mathrm{b}$ & basal friction coefficient between a particle and the bottom plate (-)\\
$\xi$ & displacement in the tangential direction (m)\\
$\varrho$ & component of the elliptic coordinate (-)\\
$\sigma_{\varrho\vartheta}$ & $\alpha\beta$ component of the stress (N/m$^2$)\\
$\widetilde{\sigma}_{\varrho\vartheta}$ & normalized stress\\
$\widetilde{\Sigma}_{\rm surf}$ & vector of the Fourier components on the surface of the intruder (-)\\
$\varphi$ & area fraction of the system (-)\\
$\phi$ & velocity potential function (m$^2$/s)\\
$\chi$ & Airy stress function (N)\\
$\psi$ & stream function of the perfect fluid (m$^2$/s)\\
$\Psi$ & stream function of the viscous fluid (m$^2$/s)
\end{tabular}

%%%%%%%%%%%%%%%%%%%%%%%%%%%%%%%%%%%

\end{document}